\documentclass[structabstract]{aa}  

\usepackage{graphicx}
\usepackage[]{natbib}
\usepackage{txfonts}
\newcommand{\alfa}{$\alpha$}
\newcommand{\alfaFe}{[$\alpha$/Fe]}
\newcommand{\meta}{[Fe/H]}
\newcommand{\T}{$T_{\rm eff}$}
\newcommand{\g}{log($g$)}
\newcommand{\Vmicro}{$V_{\rm micro}$}
\newcommand{\mic}{$\mu {\rm m}$}
\newcommand{\Tuc}{47~Tuc}
\newcommand{\NGC}{NGC~6397}
\begin{document}
   \title{Fluorine abundances and the puzzle of globular cluster chemical history
\thanks{Based on observations collected with the CRIRES spectrograph at the VLT/UT1 Antu telescope (Paranal Observatory, ESO, Chile), Programs 081.D-0276 \& 083.D-0416.}
}

   \author{P. de Laverny\inst{1}
          \and
          A. Recio-Blanco\inst{1}
          }

   \institute{Laboratoire Lagrange (UMR7293), Universit\'e de Nice Sophia Antipolis, CNRS, Observatoire de la C\^ote d'Azur, BP 4229,
 F-06304 Nice cedex 4, France\\
              \email{laverny@oca.eu}
       }

   \date{Received; accepted}

   \abstract
    {}   
    {A few studies have already shown that the abundance of fluorine in a few Galactic globular clusters is 
strongly varying from star-to-star. These unexpected chemical properties are an additional confirmation of the
chemical inhomogeneities already found in several globular clusters, and probably caused by the first generations
of stars formed in these systems. The aim of this article is to complement
our understanding of the fluorine behaviour in globular cluster stars and to look for new
constraints on the formation histories of their multiple stellar populations.}
    {We have collected near-IR spectra with VLT/CRIRES of 15 Red Giant Branch stars belonging to four 
different globular clusters
spanning a wide range of metallicity: \Tuc , M~4, \NGC \ and M~30. 
We have estimated the fluorine abundance of these targets from the
spectral synthesis of one HF line at 2.336$\mu$m. Sodium and iron abundances have also been determined from two individual lines of FeI and NaI found in the same spectra.}
    {No anticorrelation between fluorine and sodium abundances are found
for the most metal-rich cluster of the sample (\Tuc). In this cluster, RGB stars indeed exhibit rather small differences in [F/Fe] unlike the larger
ones found for the [Na/Fe] ratios. 
This reveals a rather inhomogeneous stellar system and a complex chemical evolution history
for \Tuc . 
In M~4, one star of our study confirms the previous Na-F distribution 
reported by another group in 2005.
For the two very metal-poor globular clusters (\NGC \ and M~30), only upper limits of fluorine abundances have been
derived. We show that F abundances could be estimated (but with high uncertainty) in such 
metal-poor globular clusters with current telescopes and spectrographs only if unexpected F-rich giants are 
found and/or exceptional observational conditions are met.}
  {The distribution of the fluorine and sodium abundances in globular clusters reveal that their RGB members
seem to belong to two well-separated regions. 
All the RGB stars analysed so far in the different globular clusters are indeed
found to be either F-rich Na-poor or F-poor Na-rich. Such well-separated bimodal
regimes are consistent with the separate formation episodes suspected in most galactic globular clusters.} 

   \keywords{Galaxy: globular clusters - Stars: abundances - globular clusters: individual: 47 Tuc, M~4, NGC~6397, M~30  } 

   \maketitle
%

\section{Introduction}

It is now well admitted that galactic globular clusters (GC) have
a formation history that is much more complex than previouly thought.
Several signatures of successive stellar populations have
indeed been reported in most GCs \citep[see][for an extensive review]{Gratton12}.
However, the different episodes of stellar formation in these clusters are still not
well understood. One possible way to disentangle this problem is to get
informations on the chemical evolution history of each GC.
The chemical signatures of the previous populations can
indeed be searched at the surface of the present stars in these clusters.
Up to now, two main ranges of mass for the first-generation polluting stars
in GCs have been favoured: intermediate-mass on the Asymptotic Giant Branch (AGB) or fast-rotating massive stars
on the main sequence \citep[see. for instance,][]{Charbonnel13, Renzini13}.

A very interesting chemical element for getting a better view of GC multiple stellar populations is fluorine.
Indeed, although the exact origin of this chemical specy in the Universe is still an open question,
there are only two main ranges of stellar masses in which it is suspected to be produced:
massive stars (type II supernovae and/or Wolf-Rayet stars) or low-mass stars
on the AGB \citep[see][for a more detailed description of these production
mechanisms]{Recio12}. However, from an observational point of view, 
there is direct evidence of fluorine production only in low-mass AGB stars \citep{Jorissen92, Abia09, Abia10, Recio12}.
In contrast, two main types of stars are known to destroy fluorine:
fast-rotating massive stars \citep{Prantzos07} and
intermediate-mass AGBs that experience Hot-Bottom Burning process, which is also a phase
of sodium production. Therefore, exploring the F and Na content in GC may help
disentangle the formation history of these clusters.

Unfortunately, the abundance of fluorine in globular cluster
stars has still not been explored or understood very well.
Up to now, F-abundances have been reported
for very few stars found in only four different GCs.
The first fluorine abundances in GC
stars were determined by \citet{Cunha03} for two Red Giant
Stars (RGB) of $\omega$~Cen. These stars belong to the metal-rich
component of this GC (their metallicity is [Fe/H] = -0.9 and -1.2~dex), and actually, only an upper
limit of the F-abundance has been derived for one of them.
The other RGB target in $\omega$~Cen exhibits a moderate fluorine content ([F/Fe]=-0.2~dex).
Then,  \citet{Smith05} published the F-content of seven giant stars
of the GC M~4, which mean metallicity is [Fe/H]=-1.18~dex \citep{Carretta09}.
This fundamental work has shown that the fluorine abundances strongly vary from
star-to-star ([F/Fe] is varying from $\sim$-0.75 to $\sim$-0.1~dex) and that they are
anticorrelated with sodium abundances. This leads \citet{Smith05} to suggest that a link may
exist between the F-abundances and the chemical pollutions
caused by previous stellar generations in M~4 in which intermediate-mass AGB stars
might have played a major role. At similar metallicity, 
\citet{Yong08} have studied the fluorine content in five RGBs of the GC NGC~6712 ([Fe/H]=-1.0~dex).
They again found a large scatter in the F abundances: [F/Fe] is varying from -1.0 to -0.17~dex.
The analysed stars can be grouped into two different regimes: two stars are found to be 
F-poor and three are F-rich. Such a picture could be interpreted, as in M~4, with 
the yields of intermediate-mass AGB and/or those of massive first-generation stars.
Finally, fluorine abundances have also been studied in the more metal-poor massive GC
M~22 with a mean metallicity [Fe/H]$\sim$-1.7~dex and a dispersion around 0.2~dex. 
\citet{Alves-Brito12} derived F-abundances in five RGB stars and
upper limits in two other stars of this cluster. No correlation between F and O or Na  
abundances is seen by these authors. This conclusion has been ruled out
by \citet{Dorazi13}, who tried to derive fluorine abundances in six RGB of M~22
\citep[four of them in common with][]{Alves-Brito12}. It is shown
that the telluric substraction inherent to any K-band spectroscopy 
may partly explain the largest differences in the F-abundances derived by both groups.
However, these rederived F abundances in M~22 can still be questioned (see our discussion
on F-abundance determination in metal-poor GC stars in Sect~\ref{GCPoor}),
and a deeper analysis of these spectra reveal that no
definitive conclusions on the F and Na distributions in M~22 stars
can be drawn \citep[see][]{deLaverny13}.

To provide new constraints
about the nature of the first-generation stars that played a major role in
the chemical evolution of globular clusters, we present in this work the fluorine abundances (or upper limits)
of 15 RGB stars belonging to four GC spanning a
wide range of metallicity (from [Fe/H]=-0.76 to -2.33~dex).
This paper is organised as follows:
Section~\ref{Sect:Obs} is devoted to the description of the observations, the selection of the
targets, the determination of their stellar parameters, and chemical analysis.
We then present
our abundance determinations in the most metal-rich clusters studied in Sect.~\ref{Sect:discu}. Our results are discussed in Sect.~4
with a particular
emphasis on the limitations of the F abundance determinations in metal-poor GC. 
Finally, Sect.~\ref{conclu} summarizes our conclusions.

\section{Observations and chemical analysis}
\label{Sect:Obs}
\subsection{CRIRES observations and targets selection}
The spectra were collected with the VLT/CRIRES spectrograph during two nights
in July 2008 (Program 081.D-0276) and in service mode in summer 2009 (Program 083.D-0416).
We adopted the CRIRES standard setup {\it WLEN.ID=24/-1/i} that allows the observation
of the HF(1-0)~R9 line at $\sim$2.336~\mic \ on which our abundance analysis is based.

The selected targets are giant stars located close to the tip (or slightly below) of the RGB of four globular clusters. 
Bright RGB stars have been favoured since the HF line appears
stronger in cooler stars. We also observed slightly less bright stars in order to explore
the possibility of detecting the F line in hotter but possibly more F-enriched stars and to avoid the possible confusion
between RGB and AGB stars where fluorine could be synthetized.
To explore the variation in fluorine between different chemical
environments, we also selected GC spanning a wide range of metallicities:
metal-rich (\Tuc), intermediate metallicity 
(M~4), and very metal-poor GC (\NGC \ and M~30). 
The RGB stars of these clusters have been selected from the colour-magnitude diagrams of
\citet{Rosenberg00a} for M~4 and \NGC , \citet{Momany04} for M~30 and 
Momany (private communication) for \Tuc. These works provide the $BVI$ photometry of the selected targets,
and we also collected their 2MASS $JHK$ magnitudes. The observed stars are listed in Table.~\ref{Tab:param} with
the naming convention of their visible photometry together with their $K$-band magnitude. The GC membership
of the selected targets was confirmed with the radial velocity derived from the CRIRES spectra.
Together with these RGB stars and in order to remove the lines of the Earth's atmosphere
superimposed on the stellar spectra, several telluric standards were observed at the same airmass 
as the science targets, just before and after them. The selected standards were featureless
hot OB stars. The spectra were reduced with the
ESO/CRIRES pipeline, and standard IRAF procedures
were used to remove the
telluric contributions.

\subsection{Stellar atmospheric parameters}
We first assigned to every star of each cluster the mean cluster metallicty estimated by \cite{Carretta09},
together with an \alfa -elements enhancement with respect to iron (\alfaFe) as found in most galactic GC stars for a given \meta.
Then, the effective temperature of each star has been computed as the mean of the temperatures estimated from the (V-J) and (V-K) 
colours following the
calibration for giant stars of \citet[their Eq.2]{Ramirez05}. 
The differences between these two temperature estimates are
always less than $\sim$60~K and less than $\sim$15~K for 75\% of the sample.
When computing these \T \ values, we adopted the extinction from the GC foreground reddening reported by \citet[revised in 2010]{Harris96} 
and from the relations given in \cite{Cardelli89}.
We point out that the differential reddening present in the direction of the GC M~4 does not strongly affect our \T \ estimates
since we mostly rely on the IR photometry of the targets.
The two estimates of \T \ for the three targets of M~4 indeed vary by less than $\pm$40~K.
The surface gravities were calculated from the derived effective temperatures, the dereddened magnitudes,
the bolometric correction estimated from the relation provided by \cite{Alonso99}, and the distance (Harris catalogue), and we adopted a typical mass of 0.8~$M_{\odot}$ for the RGB stars in our sample. A slightly heavier mass of 
0.9~$M_{\odot}$ would increase \g \, by only 0.05~dex, and would have a negligible effect
on the derived abundances (see below the discussion on the error analysis). The error on the other
parameters used for the surface gravity estimates can also be neglected since their impact on the derived
\g \ is very weak.
Finally, the microturbulent velocity was estimated from the empirical relation for metal-poor 
giant stars derived by \cite{Pilachowski96}.

We have searched the literature for checking our atmospheric parameter estimation. 
One of our targets (\#51362 in \NGC) has already been studied 
in detail by \citet{Lind11}. The differences between their adopted atmospheric parameters
and ours are only +87~K, 0.02~dex, and 0.01km/s for \T , \g \ , and \Vmicro, repectively. Moreover, our brightest target in the GC M~4 (\#45330) can be compared to the faintest one (\#$L3413$) of 
\citet{Smith05}, which both have very close \T \ (within 30~K). Although the stellar parameters are derived from
different methods (spectroscopic versus photometric), they do agree: within 0.1~dex for the surface gravity and 0.35~km/s for
\Vmicro. These different checks confirm the validity of our procedure in estimating the stellar atmospheric parameters 
adopted for the chemical analysis. All these parameters are reported in Table~\ref{Tab:param}.
\begin{table*}[t]
\caption{Atmospheric parameters and derived chemical abundances for stars in our sample.}
\label{Tab:param}
\begin{tabular}{c c c r r c c c r r r r}
\hline\hline
& & & & & & & & & & & \\
 & \meta \tablefootmark{a} & \alfaFe \tablefootmark{a} & Star & K(mag) & \T (K) &  \g \, (cm.s$^{-2}$) & \Vmicro (km/s) & [Fe/H]\tablefootmark{c} & [Na/Fe] & $A$(F)\tablefootmark{b} & [F/Fe] \\
& & & & & & & & & & & \\
\hline 
& & & & & & & & & & & \\
Arcturus & -0.52 & +0.31 & & & 4286 & 1.66 & 1.74 & -0.50 & +0.23\tablefootmark{c} & 4.16 & +0.10\tablefootmark{c} \\
& & & & & & & & & & & \\
\hline 
& & & & & & & & & & & \\
\Tuc & -0.76  & +0.3 & \#41806 &  7.36 &  3500  &  0.1 &  2.6 & -0.98 & +0.01\tablefootmark{c} & 3.15 & -0.43\tablefootmark{c} \\
      &        &      & \#68261 &  7.93 &  3800  &  0.6 &  2.3 & -0.90 & +0.23\tablefootmark{c} & 3.4 & -0.26\tablefootmark{c} \\ 
      &        &      & \#56265 &  8.49 &  3880  &  0.8 &  2.3 & -0.80 & +0.13\tablefootmark{c} & 3.5 & -0.26\tablefootmark{c} \\
      &        &      & \#68039 &  8.78 &  4015  &  1.0 &  2.1 & -0.90 & +0.53\tablefootmark{c} & $<$3.5 & $<$-0.16\tablefootmark{c} \\
      &        &      & \#38841 &  8.93 &  4015  &  1.0 &  2.1 & -0.90 & +0.23\tablefootmark{c} & 3.5 & -0.16\tablefootmark{c} \\
      &        &      & \#86622 &  8.99 &  4020  &  1.0 &  2.1 & -0.95 & +0.43\tablefootmark{c} & 3.6 & -0.11\tablefootmark{c} \\
& & & & & & & & & & & \\
M~4      & -1.18& +0.4 & \#45330  &  8.16 &  4145 &   1.3 &  2.0 & -1.10 & +0.03\tablefootmark{c}& 3.4   & -0.06\tablefootmark{c} \\
        &      &      & \#38229  &  9.28 &  4360 &   1.8 &  1.9 &       & +0.31\tablefootmark{d}& $<$4.0& $<$+0.62\tablefootmark{d} \\
        &      &      & \#39797  & 10.07 &  4375 &   2.1 &  1.8 &       & +0.11\tablefootmark{d}& $<$4.0& $<$+0.62\tablefootmark{d} \\ 
& & & & & & & & & & & \\
\NGC    & -1.99& +0.4 & \#73589 &  7.89 &  4490 &   1.2 &  1.7 & $<$-1.5 & +0.12\tablefootmark{d}     & $<$3.0 & $<$+0.43\tablefootmark{d} \\ 
        &      &      & \#73212 &  8.43 &  4565 &   1.4 &  1.7 & $<$-1.5 & $<$-0.08\tablefootmark{d} & $<$3.3 & $<$+0.73\tablefootmark{d} \\ 
        &      &      & \#51362 &  8.52 &  4640 &   1.5 &  1.6 & $<$-1.5 & +0.02\tablefootmark{d}     & $<$3.5 & $<$+0.93\tablefootmark{d} \\ 
        &      &      & \#52830 &  8.54 &  4630 &   1.5 &  1.6 &         & +0.12\tablefootmark{d}     & $<$3.7 & $<$+1.13\tablefootmark{d} \\ 
& & & & & & & & & & & \\
M~30   & -2.33  & +0.4 & \#3998  &  8.86 &  4150 &   0.5 &  2.0 &         & +0.16\tablefootmark{d} & $<$2.6 & $<$+0.37\tablefootmark{d} \\
      &        &      & \#7640  &  9.77 &  4445 &   0.9 &  1.8 &         & +0.46\tablefootmark{d} & $<$3.2 & $<$+0.97\tablefootmark{d} \\
& & & & & & & & & & & \\
\hline
& & & & & & & & & & & \\
\end{tabular}
\tablefoot{
\tablefoottext{a}{These columns refer to the mean \meta \ ratios of each cluster \citep{Carretta09} and 
the corresponding \alfaFe \, ratios as seen in most galactic stars for a given \meta . For Arcturus, the stellar parameters and 
\alfaFe \ are from \citet{Ramirez05}.} \\
\tablefoottext{b}{For the chemical abundances, we use the classical relation $A$(F) = log [$n$(F)/$n$(H)] + 12. The adopted F solar abundance
is 4.56.}\\
\tablefoottext{c}{These chemical ratios are computed with the iron abundance measured with the FeI line at 2.3308\mic .  The adopted 
Na and Fe solar abundances are 6.17 and 7.45, respectively.}\\
\tablefoottext{d}{These chemical ratios are computed with the mean iron abundance of each cluster  given in col.2.}\\
}
\end{table*}

\subsection{Chemical analysis}
\begin{figure}[t]
\includegraphics[width=8.5cm,height=8cm]{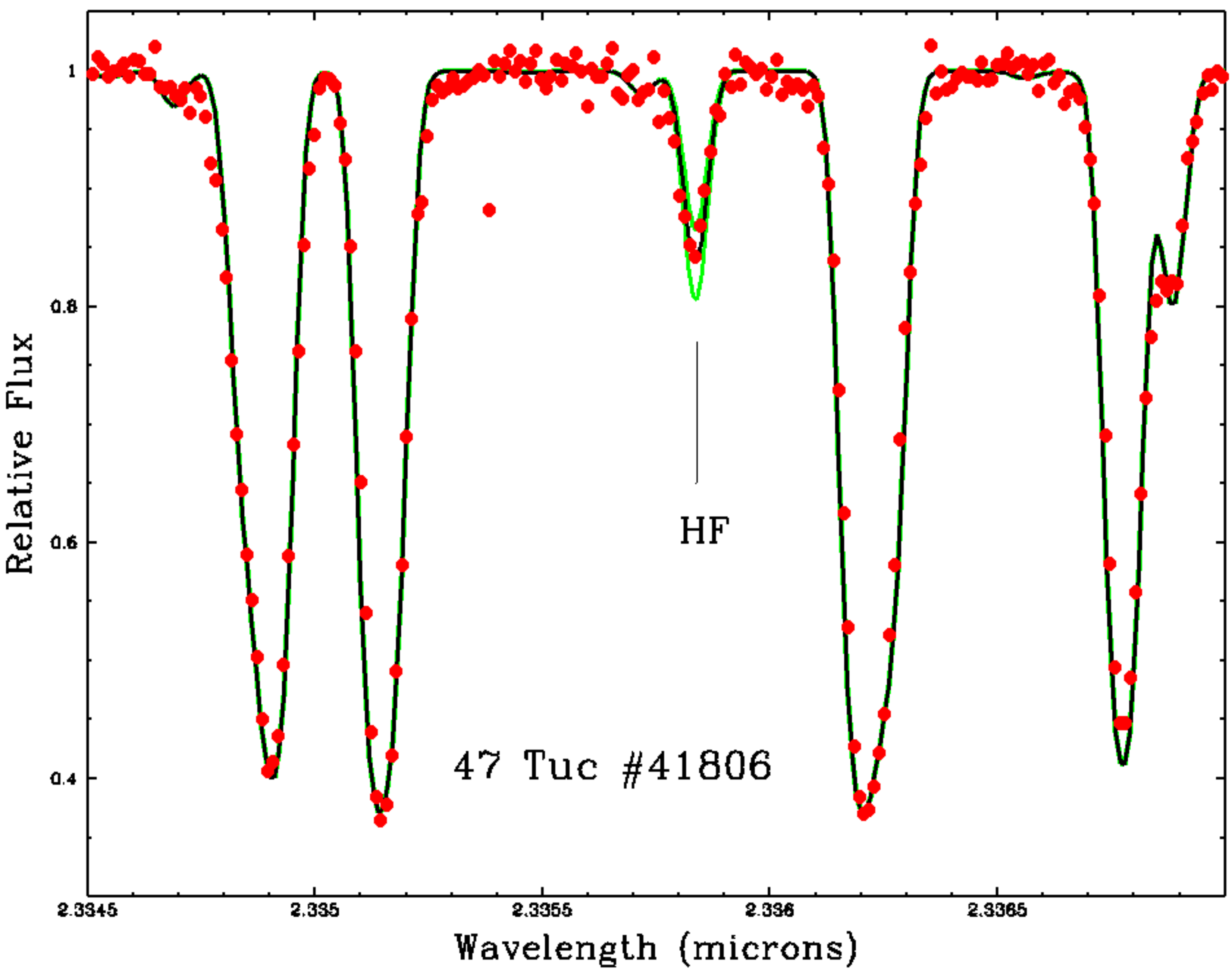}  
\caption{Observed (filled red circles) and synthetic spectra (continuous lines) around the HF(1-0)~R9 line for the star \Tuc \ \#41806.
The best fit (black line) corresponds to the fluorine abundance reported in Table~1 and synthetic
spectra computed with $\pm$0.1~dex changes in the F abundance are superimposed (green
lines).}
\label{FigSpec}
\end{figure}

The fluorine abundances were derived from the unblended HF(1–0)~R9 line at 2.3358~\mic. We have shown in
\cite{Abia09} that this line is the best F indicator in this spectral range. This domain also contains
several CO lines, together with a non-blended NaI line at 2.3379~\mic \ and a FeI line at 2.3308~\mic \
that have been used for this chemical analysis.
We adopt the same linelist as in our previous works \citep[see][]{Abia09,Recio12}.
The theoretical spectra were computed with the TurboSpectrum code \citep[and further improvements by Plez]{Alvarez98}. Spherical MARCS model atmospheres \citep{Gustafsson08} with standard chemical composition (i.e. with \alfa \, enhancements typical of the low-metallic clusters under study), a microturbulent parameter of 2~km/s and
a mass of 1~$M_{\odot}$ were adopted. For the different stars under study (see Table~\ref{Tab:param}), interpolations at their estimated stellar atmospheric parameters were performed. Solar abundances are those adopted
by \citet{Gustafsson08}. We also assumed a carbon isotopic ratio $^{12}$C/$^{13}$C=5 for all our targets, and we checked that this
assumption does not affect our derived abundances.
Finally, the synthetic spectra were broadened by convolution with a
Gaussian profile of FWHM=6~km/s to match the observed line widths. The derived F, Na, and Fe abundances (or the estimated upper limits) are
reported in Table~1, and we show in Fig.~\ref{FigSpec} an example of an observed and synthetic spectrum for the 
brightest target of \Tuc .

This procedure for the chemical analysis was firstly checked by fitting the high-resolution IR spectrum of Arcturus \citep{Hinkle95} 
and adopting the stellar parameters and abundances of \citet{Ramirez05}. Our derived abundances
are reported in Table~\ref{Tab:param}. These IR NaI and FeI abundances differ by  +0.07~dex and -0.02~dex,
respectively, with respect to those of \citet{Ramirez05} that were derived from the optical domain (and considering
that they adopted a solar iron abundance of 7.50, whereas we adopt 7.45 in the present work).
We also found an abundance of fluorine in Arcturus of 4.16 (i.e. [F/Fe] = +0.10), in total agreement with our previous determination
\citep{Abia09}, although the adopted atmospheric parameters slightly differ since in the present
work we favoured the Arcturus stellar parameters of \citet{Ramirez05}.
Moreover, we also checked our abundance determinations with the Na-rich second generation star \#51362 in \NGC \ studied 
by \citet{Lind11}. They report a sodium abundance $A$(Na)=4.23 that is
very close to the value we measured $A$(Na)=4.2. Furthermore, their
metallicity \meta \ lies between -2.06 and -2.10, and it is consistent with
the upper limit we derived ($<$-1.5).

On the other hand, we point out that the estimated chemical abundances reported in Table~\ref{Tab:param} reveal that the derived metallicity for each star of \Tuc \
is slightly more metal-poor than expected from the mean metallicity of this cluster (adopted from \citealt{Carretta09}).
Our mean [Fe/H] is indeed about 0.14~dex lower than the \citet{Carretta09} one, i.e. a difference
larger than our estimated
uncertainty that cannot be explained by the different adopted solar reference values. 
This could be explained by the fact that our [Fe/H] derivations rely on 
only one FeI line and could be less robust because of telluric line contaminations that could
alter the line profile and the continuous position. 
However, even if some iron abundance scatter is found from star-to-star
within the present sample, this scatter is smaller than the estimated error and can thus be 
attributed entirely to the measurement uncertainties. The analysed iron line is therefore a rather
good indicator of the stellar metallicity. This is also confirmed by the correct metallicity
derived for Arcturus and by the rather good
agreement found between the iron abundance derived by us in one star of M~4 and the \citet{Carretta09} 
mean value. Therefore, the [Fe/H] values derived from this line
have been adopted in the present study when available.

Finally, we estimated that the typical errors on the derived abundances induced by uncertainties
on the stellar atmospheric parameters are 
around $\pm$0.2~dex, $\pm$ 0.1dex, and $\pm$0.1dex for fluorine, sodium, and iron, respectively.
These estimates were computed as follows: for the target
\Tuc \ \#68261 of Table~1 (which is typical of our sample), we re-estimated its abundances by varying its stellar parameters by
$\Delta$\T \ of $\pm$100~K, $\Delta$\g \ of $\pm$0.2~dex,
$\Delta$[Fe/H] \ of $\pm$0.1~dex, and $\Delta$\Vmicro \ of $\pm$0.5~km/s.
These different contributions were
then summed in quadrature, together with the error in the fit (around $\pm$0.05~dex)
dominated by the continuum placement and the telluric residuals. We point out that
other possible systematic sources of uncertainties (as log~gf uncertainties, for instance) 
are not taken into account in this error budget. 

\section{Fluorine abundances in \Tuc \ and M~4}
\label{Sect:discu}
We present in this section the estimated fluorine abundances in \Tuc \ and M~4 stars
(i.e. the most metal-rich clusters of the present study). 
We refer to the next section for the discussion on the F abundance upper limits
derived for the two other studied GCs that are much more metal-poor (\NGC \ and M~30).

\subsection{\Tuc}
\begin{figure}[t]
\includegraphics[width=8.5cm,height=8cm]{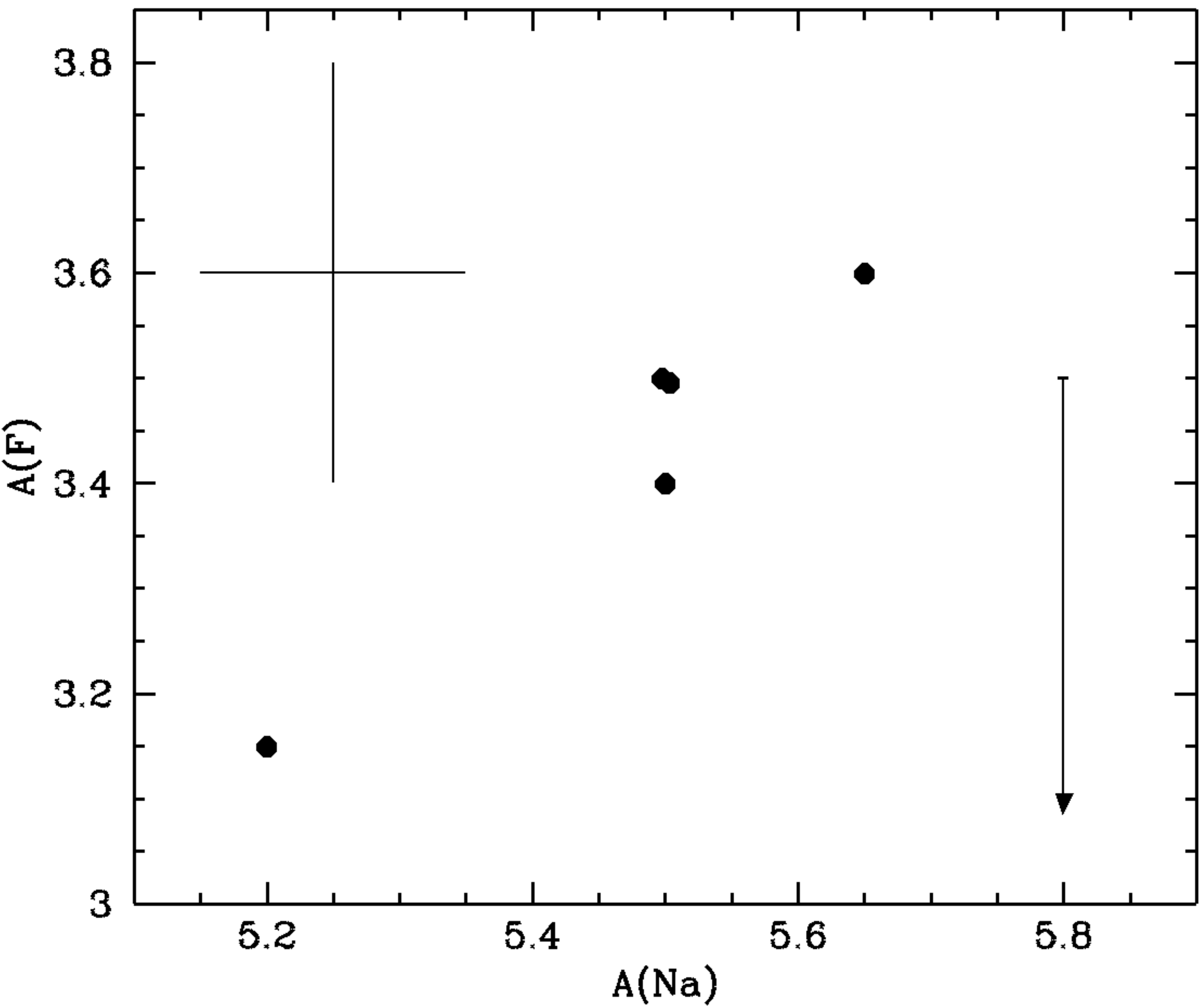}  
\caption{Fluorine abundances with respect to sodium abundances for the RGB stars of \Tuc.
For one star (\#68039), only an upper limit for fluorine has been derived. It is indicated with an
arrow oriented towards low fluorine abundances.
Within the abundance uncertainties, no (anti)-correlation are detected between fluorine and sodium.}
\label{FigNaF}
\end{figure}

For the fluorine content of RGB stars in \Tuc , we illustrate in Fig.~\ref{FigNaF} the derived 
F abundances versus the sodium ones. 
Although the error bars of each determination are large, we detect, as in all
the other clusters studied so far, a large scatter in the F and Na abundances.
The amplitude of these abundance variations exceeds the measurement uncertainties.
The most fluorine-rich star is found to be enriched by 0.45~dex with respect to the most
F-poor one and their Na-content also differs by almost the same amount
(but their iron content are quasi-identical). On the other hand, 
the range of the sodium abundances for these six stars of \Tuc \ is even broader and spans about 0.6~dex.
\Tuc \ is thus a rather inhomogeneous cluster when one examines its F and Na chemical properties.

However, from this Fig.~\ref{FigNaF},
no clear (anti-)correlation between the fluorine and sodium chemical abundances is seen
with the present sample, unlike what has been reported in the other galactic globular cluster M~4
(see Fig.~\ref{FigNaFe_FFe} and next subsection).
Such an anticorrelation in \Tuc \ is mostly ruled out by two stars of the sample
(\#41806 and \#86622).
First, the RGB star \#41806 is indeed found to be very poor in F and Na (see its spectrum
in Fig.~\ref{FigSpec}).
We have carefully checked this spectrum, the telluric contaminations, and 
the abundances derived for this star, and we are confident of them.
Furthermore, although one possible source of uncertainty could be the estimation of the effective 
temperature, we have checked that adopting a different \T \ would not reveal any Na-F anticorrelation
in \Tuc .
For instance, a hotter \T \ would lead to a more enriched star in both fluorine and sodium, thus 
ruling out anyway a possible anti-correlation.
Secondly, the other star that could be incompatible with a Na-F anticorrelation is the target \#86622.
This RGB is indeed not only Na-rich but also very enriched in fluorine. It is actually the most F-rich
star we have observed in \Tuc . 

With the present sample, we are therefore confident that no anticorrelation between F and Na is detected in \Tuc .
Increasing the size of the sample would perhaps help to confirm or not this conclusion. However, 
if an anticorrelation were found in the future, 
two stars in our study (representing one third of our total sample) 
would be rather exotic, and their
specific Na and F properties should be understood.

\subsection{M~4}
For this globular cluster, our study adds a new point in the F-Na relation found
by \citet[see their Fig.~4]{Smith05}. This new measurement (the RGB  star \#45330) perfectly confirms their finding, adding one new F-rich Na-poor star in this figure. This star 
could even be the most F-rich RGB ever studied
in any GC (in terms of [F/Fe] ratios) with [F/Fe] almost solar. 
Moreover, it is interesting to note that this target \#45330 has very close 
atmospheric parameters and almost the same F and Na abundances (within 0.1~dex)
as the faintest star analysed by \citet[their star \#L3413]{Smith05}.

The two other RGB stars of M~4 analysed in the present study appear to be rather Na-poor,
but we have been able
to derive only upper limits of their F content\footnote{For clarity reasons, these two stars are not
shown in the Fig.~\ref{FigNaFe_FFe} described in Subsect.~4.1}. 
This is probably caused by the fact that these two stars
are not located at the tip of the RGB and are thus rather hot, leading to too weak an HF line
to be detected well. 
Indeed, the only M~4 star for which we have been able to derive a fluorine abundance (\#45330) is
about one and two magnitudes brighter in the K-band than these two M~4 stars with estimated F upper limits, respectively. 
It is also cooler by more than 200~K than these two fainter targets. 

\section{Discussion}
\subsection{Fluorine abundances distribution in metal-rich or intermediate-metallicity globular clusters}
\begin{figure}[t]
\includegraphics[width=8.5cm,height=8cm]{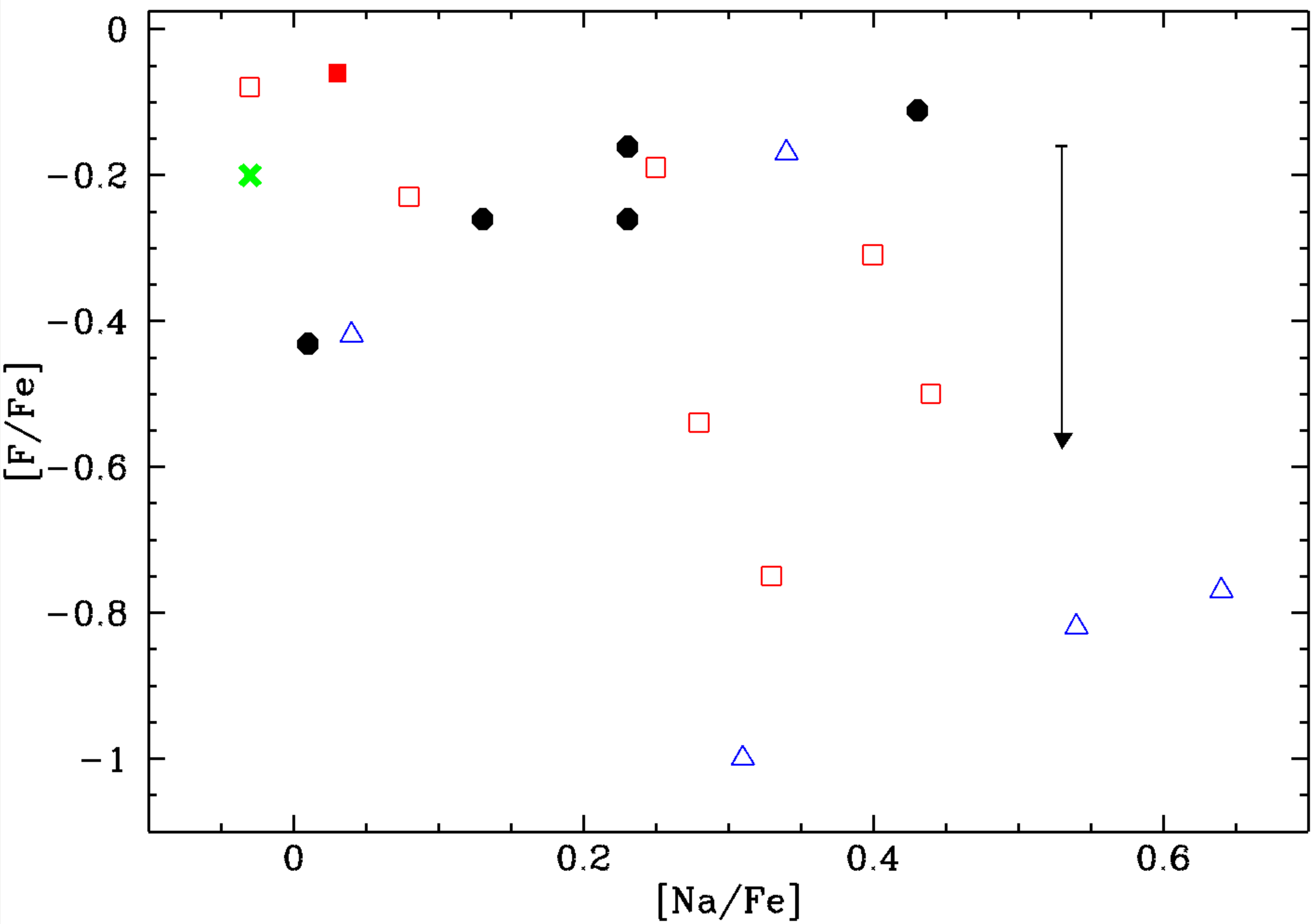}  
\caption{[F/Fe] versus [Na/Fe] ratios for RGB stars found in metal-rich or intermediate-metallicity globular clusters:
\Tuc \ (this work, filled black circles), M~4 \ (this work, red filled square and 
\citealt[red empty squares]{Smith05}), NGC~6712 \citep[blue empty triangles]{Yong08}
and $\omega$~Cen \citep[green cross]{Cunha03}. 
The RGB star of \Tuc , for which we have derived an upper limit for the fluorine 
abundance, is indicated with an
arrow oriented towards low [F/Fe] ratios.
The [F/Fe] upper limits derived for the stars belonging to the
very metal-poor clusters \NGC \ and M~30 are located well above the upper limit of this plot.
Typical errors are around $\pm$0.25~dex in both axes.}
\label{FigNaFe_FFe}
\end{figure}
 
Up to now, fluorine abundances have been safely derived in RGB stars belonging
to the metal-rich GC \Tuc \ (this work) and $\omega$~Cen \citep{Cunha03}
and to the intermediate-metallicity GC M~4 (this work and, mainly, \citealt{Smith05}) and
NGC~6712 \citep{Yong08}\footnote{See the introduction of the present article for a summary
of these works.}. The mean metallicity of these GCs
is found to lie between -0.76 and -1.2~dex, and
we refer to Sect.~\ref{GCPoor} for the discussion on the fluorine abundance estimates
in more metal-poor clusters.

First, we compare in Fig.~\ref{FigNaFe_FFe} the [F/Fe] and [Na/Fe] ratios in these four rather metal-rich GC.
In this plot, for the unique RGB star of $\omega$~Cen with a well defined fluorine abundance i.e. not an upper limit \citep{Cunha03}, we adopt its sodium abundance derived by \citet{Smith00}. 
Regarding a possible anticorrelation between fluorine and sodium in GC, it could be 
seen in this Fig.~\ref{FigNaFe_FFe} only for M~4, but the large errors on the abundance ratios could
hide different behaviour. Indeed, even in this GC, large F-abundance variations for almost
constant Na-abundances are present. For instance, one can find three stars (over a sample
of only 8 stars) having almost identical Na-abundances ([Na/Fe] $\sim$ 0.25-0.3~dex), but their [F/Fe] ratios
differ by more than 0.6~dex, well beyond the error bars.
Such a spread in F and not in Na could reveal a rather inhomogeneous chemical history within M~4, 
particularly regarding the nature of
its F-polluters. In any case, such a figure is not consistent with a classical anti-correlation
between fluorine and sodium but, owing to the rather large abundance uncertainties, this cannot be  concluded definitively.

For the three other clusters, \Tuc , NGC~6712 (in which
five stars have an estimated F abundance in each of them),
and $\omega$~Cen (only one star),
the stars are located into two separated regions of the diagram: {\it F-rich Na-poor} and 
{\it F-poor Na-rich}. From their Na content, it can be claimed
that these two groups correspond to the first- and second-generation stars
of these GCs, respectively.
Moreover, no smooth variation is seen between these two regimes. 
Indeed, one finds two stars
(\#V10 and \#LM10 of \citealt{Yong08})
in NGC~6712 with almost identical [Na/Fe] but differing by 0.8~dex in [F/Fe].
It is difficult to claim that any continuous anticorrelation does exist in NGC~6712 when
one looks at these two stars. Finally, our RGB targets in \Tuc \ are all located
in the {\it F-rich Na-poor} part of Fig.~\ref{FigNaFe_FFe}. It would be nice to see
if future studies will report any {\it F-poor Na-rich} second-generation stars in this cluster (our target
star \#68039 with only an upper limit of its fluorine abundance could be one of these).
We recall that three distinct populations have been photometrically found by
\citet{Milone12} in \Tuc . Therefore, it is highly probable that {\it F-poor Na-rich} ones are
present in this cluster.

In summary, with
the present situation illustrated in Fig.~\ref{FigNaFe_FFe}, it can only be said that, except perhaps for M~4, 
RGB stars in GCs belong to at least two rather well-separated groups in an F-Na diagram: they are
either {\it F-rich Na-poor} or
{\it F-poor Na-rich}. Actually, such a dichotomic view cannot be excluded for M~4 if one considers
the errors in the F and Na abundances.
Moreover, in all these GCs, 
the spread in Na-abundances for the F-rich stars (which all have almost
identical [F/Fe] around -0.2~dex) is rather
large (up to 0.5~dex).
It can therefore be concluded that the chemical evolution of fluorine 
in these clusters is probably more complex than previously claimed and that GCs have probably been formed from very
inhomogeneous media since they contain stars with very similar Na abundances but very different F-abundances
and/or rather large F-enrichment with a wide spread in Na.

Finally, it is interesting to point out that
the two separated regimes of the F-Na abundances in GC giants could be linked
to the separated sequences revealed in colour-magnitude diagrams of most GCs. Such distinct photometric
sequences can be opposed to the more continuous ones proposed by most spectroscopic studies.
However, such spectroscopic analysis suffer from quite large abundance
measurement errors that could hide a multimodal distribution \citep[see, for instance,][]{Renzini13}. 
Therefore, the two regimes
seen in Fig.~\ref{FigNaFe_FFe} are indeed consistent with different star formation episodes in globular clusters,
thus confirming the photometric description. 

\subsection{Fluorine in metal-poor globular clusters}
\label{GCPoor}
\begin{figure}[t]
\includegraphics[width=8.5cm,height=8cm]{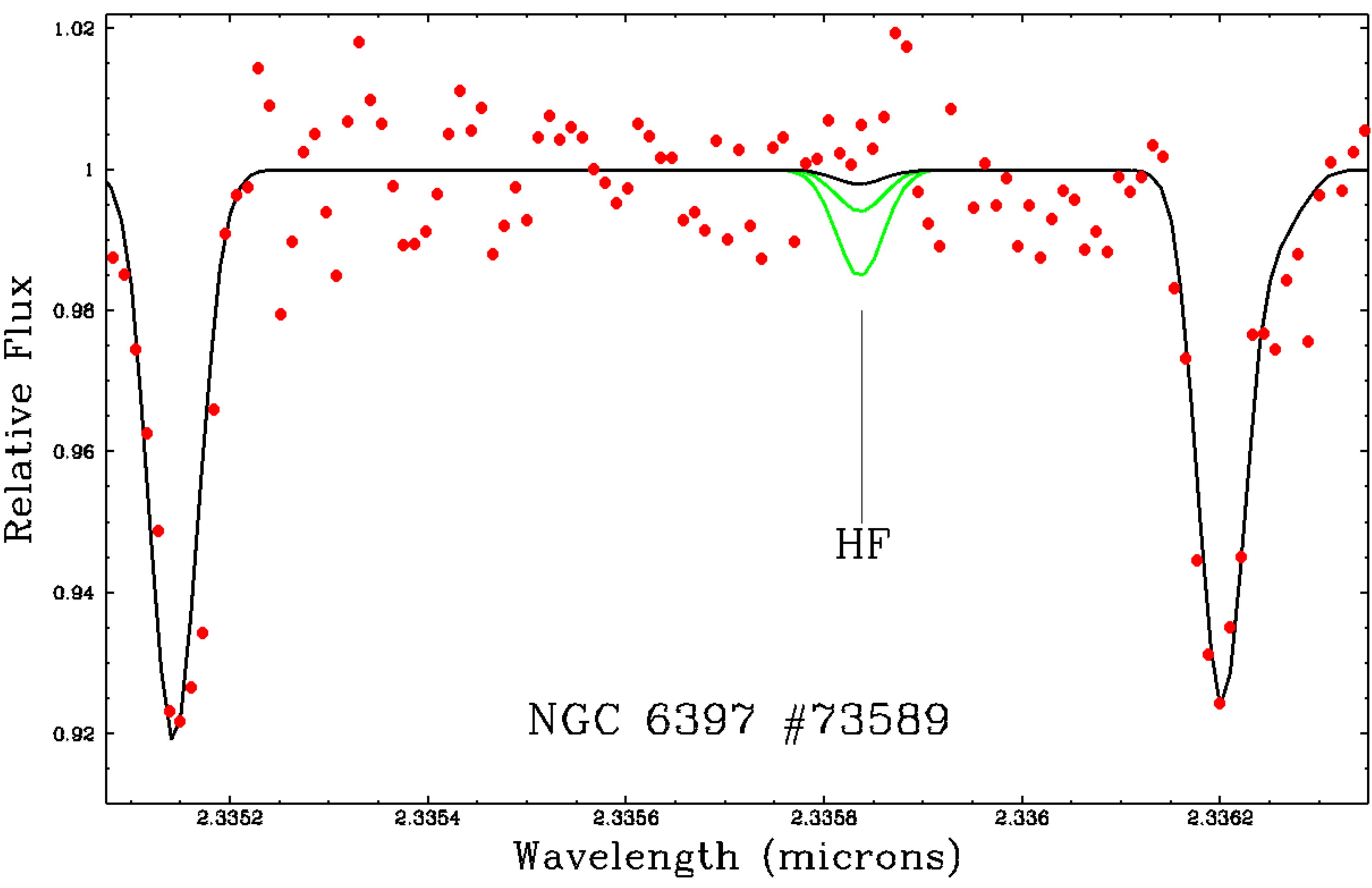}  
\caption{Observed (filled red circles) and synthetic spectra (continuous lines) around the HF(1-0)~R9 line for the star \NGC \ \#73589.
The black line corresponds to a synthetic spectrum computed for the atmospheric parameters of this star reported in Table~1 and 
adopting the largest fluorine abundance ([F/Fe]=-0.1) ever
measured in any GC. The green lines correspond to synthetic
HF lines for a very evolved star at the RGB tip of \NGC \ with \T = 4\,250~K and unrealistic F enrichments for a GC,
[F/Fe]=-0.1 and +0.3~dex, respectively.}
\label{FigSpec_N6397}
\end{figure}
Fluorine abundance studies in metal-poor GC ([Fe/H] $< \sim$ -1.5~dex) have only
been reported in  \NGC \ and M~30, which are extremely metal-poor ([Fe/H] $<$ -2.0~dex; present work) and in M~22 with stars having [Fe/H] between -1.6 and -1.9~dex \citep{Alves-Brito12, Dorazi13}.

We have only been able to derive upper limits of the fluorine content in the extreme metal-poor
GC \NGC \ and M~30. 
These F abundance upper limits are not shown in Fig.~\ref{FigNaFe_FFe} for clarity, since they are 
much higher than the classical [F/Fe] ratios found in more metal-rich clusters.
Such a limitation in fluorine determination in metal-poor stars is caused by the possibiliy that, at the CRIRES resolution and even with a rather optimal observing strategy for the telluric standards, the telluric
line residuals can have the same amplitude as the HF signature (about $\pm$1\% around the continumm level
in the present case). We think that even under optimal and steady weather conditions, it
is very difficult to get telluric residuals with an amplitude that is at least twice smaller.
Furthermore, metal-poor stars are known to be quite hotter than their metal-rich counterparts.
This leads to an even weaker HF line since this molecular transition appears stronger 
in cooler media, keeping the fluorine abundance constant.

This situation is illustrated in Fig.~\ref{FigSpec_N6397} where it can be seen that
the HF line could only be strong enough in a very cool giant star located at the tip
of \NGC \ and having an extreme (and probably not realistic for a GC) F-enrichment. In this figure, we adopt an effective
temperature close to the one of the coolest star ever studied in this cluster \citep[see, for instance,][]{Castilho00}.
Of course, the situation is even worse for the even more
metal-poor cluster M~30. This figure shows why it has been impossible to report any
interesting information on the F-content of these clusters that are too metal-poor.

Futhermore, this problem could be even more complex due to unfortunate radial velocity shifts between the telluric 
standard and the GC target spectra. Indeed, in some cases, the blue wing of the HF line could be much more
contaminated
by the telluric residuals than the red one due to the strongest telluric lines found there.
This effect leads to an even more difficult F abundance determination in such metal-poor stars. 
We think that the above problems can affect the F-abundances in M~22 reported by \citet{Dorazi13}, 
even after their revision of the abundance estimates of \citet{Alves-Brito12}. 
Such an artefact could indeed lead to a wrong identification of the very weak HF line
whereas all the other strongest lines (CO and Na) are much less affected by such a wavelength shift
\citep[see][for a more detailed discussion]{deLaverny13}.

We therefore think that fluorine abundance studies in metal-poor giant stars ([Fe/H] $< \sim$ -1.5~dex)
could face a limitation in performing the correct analysis of the HF line \cite[except
if extremely F-enriched stars are encountered as in][]{Schuler07}. 
This could only be solved
owing to much higher SNR spectra collected under very steady weather conditions
and possibly with higher spectral resolution to get a better substraction
of the telluric signatures.

\subsection{Comparison of fluorine content in Galactic and extragalactic stellar populations}
From Fig.~\ref{FigNaFe_FFe} and with the present rather small sample of
analysed RGB stars in globular clusters, 
it can be noted that GC stars are rather weakly enriched in fluorine.
Indeed, the derived [F/Fe] ratios in these GC stars (with typical values between -0.1 and -1.0~dex)
are always much lower than the ones estimated
in other samples of Galactic\footnote{This characteristic has already been
partially discussed by \citet{Yong08} in terms of fluorine abundances in Galactic stars but not regarding
[F/Fe] ratios as in the present work.} and extragalactic stars. 

For instance, bulge stars have
[F/Fe] ratios in the range [+0.25, +0.8] except one star found with [F/Fe]=-0.27 \citep{Cunha08}.
The F abundances have also been estimated in three Halo Carbon-Enriched Metal-Poor stars that
exhibit extremely high [F/Fe] ratios \citep[from 0.6 up to 2.9~dex,][]{Schuler07, Lucatello11}.
In the Galactic disc, members of the Orion Nebula cluster have [F/Fe] within -0.05 and +0.1~dex \citep{Cunha05}
and dwarfs of the solar neighbourhood or giants as Arcturus have [F/Fe] within +0.0 and +0.55~dex \citep[this work and][]{Recio12}.
Moreover, 
these low [F/Fe] ratios in GC are also much lower than those found in
low-mass AGB carbon stars in the solar vicinity in which fluorine is seen to be synthetized and dredged-up to their
surface. These low-mass stars are characterized by a [F/Fe] ratio between -0.1 and +0.65~dex \citep{Abia09, Abia10}.
Only most of the peculiar carbon stars of J spectral type (which evolutionary origin is still unknown)
have slightly negative [F/Fe] ratios \citep[between -0.2 and +0.1~dex;][]{Abia10}, but these ratios are still higher than what is found in most galactic globular cluster
giants.
Finally, even larger fluorine contents (and probably more efficient in-situ production) are found in metal-poor
extragalactic carbon stars
\citep{Abia11}. Most of the derived [F/Fe] ratios in these low-mass extragalactic AGBs lie between 0.8 and 1.7~dex, 
whereas their iron content spans the same range of metallicity as the one of the GCs studied in the present work.
Such observed large F-enrichments are consistent with theoretical models of low-mass AGB stars that predict higher [F/Fe] ratios
at lower metallicities.

In summary, the observed difference between the [F/Fe] ratios estimated in GC stars and in other galactic and extragalactic populations appears to be large (more than a factor 10).
This is another confirmation that the chemical evolution history in GCs is quite peculiar
with respect to other populations.

\section{Summary}
\label{conclu}

The IR spectral analysis of 15 stars found in different globular clusters spanning a wide range of
metallicity (from metal-rich, [Fe/H]=-0.8, to extreme metal-poor, [Fe/H]=-2.3) has allowed us to study the fluorine
and sodium abundance distributions of these clusters.
For the most metal-rich GC of the present study (\Tuc ), we reported the first description of its 
fluorine content. Fluorine abundances were found to be almost constant 
(within error bars), whereas sodium strongly varies from star-to-star, revealing a rather inhomogeneous cluster.
No Na-F anticorrelation was thus seen in \Tuc . The only M~4 star for which a fluorine abundance
has been derived (forgetting the two other M~4 targets with F upper limits) is consistent with the F-Na
distribution already reported by \citet{Smith05}. 
In the more metal-poor clusters studied (\NGC \ and M~30), only F-abundance upper limits have been derived with upper values
that are too high to provide interesting constraints on the formation history of these
clusters. We have indeed shown  that 
fluorine abundances can be safely estimated only in metal-rich or intermediate metallicity clusters.
In very metal-poor clusters, exceptional observing conditions and/or very F-enriched stars should be met
to derive right estimate of the fluorine abundances.
We also pointed out that the chemical history of fluorine in GC seems to strongly differ
from other Galactic populations found in the disc (dwarfs of the solar neighbourhood, 
Orion nebula giants, and low-mass AGBs), in the galactic bulge and halo (RGB stars) or in extragalactic
dwarf satellites (low-mass metal-poor AGBs).

Finally, in all the clusters studied so far (in the present work or in previous ones), 
the derived fluorine and sodium abundances showed that
RGB stars seem to belong to two well-separated regimes: either 
{\it F-rich Na-poor} or {\it F-poor Na-rich}. These two groups are consistent with the
first- and second-generation stars found in most GC.
Indeed, no continuous (anti-)correlation between fluorine and sodium abundances
are clearly seen in these clusters. It is not seen in \Tuc \ and NGC~6712
and even
the anticorrelation reported by \citet{Smith05} in M~4 could be compatible with these two separate
regimes because of the errors in the abundance determinations that could blur such a figure.
Such a bimodal distribution is consistent with the separated formation episodes
in galactic globular clusters as already revealed by colour-magnitude studies
in several GCs. This scenario should be tested by future models of the
chemical evolution history of GCs. 

\bibliographystyle{aa.bst}
\bibliography{biblio}

\end{document}